\documentclass{article}

\usepackage[noadjust]{cite} 
\usepackage{arxiv}
\usepackage{cite}
\usepackage[utf8]{inputenc} 
\usepackage[T1]{fontenc}    
\usepackage{hyperref}       
\usepackage{url}            
\usepackage{booktabs}       
\usepackage{amsfonts}       
\usepackage{nicefrac}       
\usepackage[final,nopatch=footnote]{microtype}
\usepackage{lipsum}
\usepackage{graphicx}
\usepackage{subcaption}
\graphicspath{ {./images/} }
\usepackage{stmaryrd}
\usepackage{amsmath,amsthm}
\newtheorem*{definition}{Definition}
\newtheorem*{property}{Property}
\usepackage{setspace}
\usepackage{scalerel}
\usepackage{tikz}
\usetikzlibrary{svg.path}
\definecolor{orcidlogocol}{HTML}{A6CE39}
\tikzset{
	orcidlogo/.pic={
		\fill[orcidlogocol] svg{M256,128c0,70.7-57.3,128-128,128C57.3,256,0,198.7,0,128C0,57.3,57.3,0,128,0C198.7,0,256,57.3,256,128z};
		\fill[white] svg{M86.3,186.2H70.9V79.1h15.4v48.4V186.2z}
		svg{M108.9,79.1h41.6c39.6,0,57,28.3,57,53.6c0,27.5-21.5,53.6-56.8,53.6h-41.8V79.1z M124.3,172.4h24.5c34.9,0,42.9-26.5,42.9-39.7c0-21.5-13.7-39.7-43.7-39.7h-23.7V172.4z}
		svg{M88.7,56.8c0,5.5-4.5,10.1-10.1,10.1c-5.6,0-10.1-4.6-10.1-10.1c0-5.6,4.5-10.1,10.1-10.1C84.2,46.7,88.7,51.3,88.7,56.8z};
	}
}
\newcommand\orcidicon[1]{\href{https://orcid.org/#1}{\mbox{\scalerel*{
				\begin{tikzpicture}[yscale=-1,transform shape]
					\pic{orcidlogo};
				\end{tikzpicture}
			}{|}}}}
\usepackage{hyperref} 

\title{Circulating Currents in Windings:

Fundamental Property}

\author{
 Taha El Hajji \orcidicon{0000-0002-0168-8180}\\
  Department of Electrical Engineering and Automation\\
  Aalto University\\
  P.O. Box 15500, 00076 Espoo, Finland\\
  \texttt{taha.elhajji@aalto.fi} \\
   \And
 Antti Lehikoinen \orcidicon{0000-0001-8170-4146}\\
  Department of Electrical Engineering and Automation\\
  Aalto University\\
  P.O. Box 15500, 00076 Espoo, Finland\\
  Smeklab Ltd, Espoo, Finland\\
  \texttt{antti.lehikoinen@aalto.fi} \\
  \And
 Anouar Belahcen \orcidicon{0000-0003-2154-8692}\\
  Department of Electrical Engineering and Automation\\
  Aalto University\\
  P.O. Box 15500, 00076 Espoo, Finland\\
  \texttt{anouar.belahcen@aalto.fi} \\
}

\begin{document}

\maketitle
\begin{abstract}
Circulating currents in windings refer to unwanted electrical currents flowing between the parallel conductors of a winding. These currents arise due to several phenomena such as asymmetries, imperfections in the winding layout, and differences in electric potential between the parallel conductors. This effect is visible typically in windings of transformers, motors, or generators. At on-load condition, this is equivalent to having a current unevenly distributed between parallel conductors. Circulating currents have two main drawbacks: increased losses in windings and potential degradation of insulation over time. The former is an intuitive property that is widely acknowledged in the literature. This paper presents a formal proof of this fundamental property, building upon the authors' previous work and embedding it within a rigorous mathematical framework. The mathematical definition of circulating currents is provided, along with a case application in an electric machine.

\end{abstract}

\keywords{circulating currents \and windings \and high frequency \and electric machine \and transformer}

\section{Introduction}

The effect of circulating currents is defined and reviewed in \cite{Access_Taha_2024}, with a focus on the windings of electric machines. Although often overlooked in the literature, circulating currents become increasingly significant as frequency rises, making them a concern in the windings of electric machines operating at high rotational speeds \cite{Machines_Taha_2023,Aerospace_Taha_2024,HighSpeed_1CN,ICEM_Taha_2024,PhD_Taha_2023}. This issue also extends to transformers and inductors functioning at high frequencies \cite{Transformer_1,Transformer_2,Transformer_3,Transformer_4,Transformer_5,Transformer_6,Transformer_7}. Under load conditions, circulating currents in parallel conductors result in uneven current distribution, meaning that the total current is not uniformly shared among all parallel-connected conductors. Circulating currents also occur in other scenarios, where an unwanted current flow through the windings, for example, in a delta-connected three-phase stator \cite{pramodCC}. In this work, we focus specifically on circulating currents within the parallel-connected conductors of a winding. However, the definitions, properties, and proofs presented in this work can be generalized to other cases where circulating currents occur.

A fundamental property of the effect of circulating currents is that it consistently leads to higher losses in windings, thereby reducing efficiency. Additionally, this phenomenon can cause localized overheating in certain conductors carrying higher currents, potentially shortening the lifespan of the equipment. This occurs primarily because the temperature limits of the insulation class, initially designed for the average flowing current in the strands, do not account for the presence of circulating currents. The causes of circulating currents in windings can be attributed to the following key factors:

\begin{itemize}
    \item \textbf{Inductance}: When two parallel-connected conductors have different inductances, the total current is unevenly shared between them. The difference in inductance can arise due to variations in conductor's dimensions, placements (either random \cite{Placement_Random_1_RefFI_Antti,Placement_Random_2_RefFI_Antti,Placement_Random_3_PhD_Antti_2017} or pre-defined \cite{Placement_Precise_1_RefUK,Placement_Precise_2_RefFI_PhD_Jahirul_2010}), asymmetries \cite{Asymmetry_1}, and imperfections \cite{Abnormality_1}, among other factors.
    
    \item \textbf{Electric Potential}: When two parallel-connected conductors have different  potentials, a circulating current flows from the conductor with higher potential to the one with lower potential. Various technical solutions can mitigate the effects of circulating currents, such as windings transposition \cite{Transposition_1_Ref22,Transposition_2_Ref23,Transposition_3_RefCN1,Transposition_4_RefCN2,Transposition_5_RefNW,Transposition_6_RefUK,Transposition_7_RefJP,Transposition_8_RefCN}.
\end{itemize}

The models used to evaluate losses due to circulating currents are reviewed in \cite{Access_Taha_2024} 
and can be categorized into three main types:
finite element models coupled with circuit analysis \cite{FEA_1_Antti}, 
analytical models \cite{Placement_Random_1_RefFI_Antti,Transposition_1_Ref22,Transposition_2_Ref23}, 
and hybrid models \cite{Hybrid_Ref33}, 
which combine finite element results with analytical formulas. 

The focus has primarily been on windings with circular wire shapes \cite{Circular_Wire_Ref18,Circular_Wire_Ref19,Circular_Wire_Ref20}, while less attention has been given to square or rectangular wire shapes (also known as Hairpin Windings), as guidelines in \cite{Guideline_1,Guideline_2} specify conductors placement in the slots to minimize circulating currents in electric machines. However, due to imperfections \cite{Rectangular_Wire_1}, 
circulating currents may still arise in windings with rectangular or square wire shapes. Other phenomena, such as the skin effect \cite{Access_Taha_2024,ICEM_Taha_2020} and the proximity effect \cite{Access_Taha_2024,ICEM_Taha_2020,SkinProx_1_UKSE,SkinProx_2_USA_Ayman,SkinProx_3_DE,SkinProx_4_UK}, can also occur in windings, particularly at high frequency. These two effects are not covered in this paper. Authors have previously reviewed these effects in \cite{Access_Taha_2024}, along with the main models in the literature, with specific applications in electric machines.

The basic terms used in this paper are introduced in Section 2. The mathematical definition of circulating currents occurring in parallel-connected conductors is provided in Section 3. Section 4, which constitutes the primary focus of this paper, presents the proof of the fundamental property of circulating currents, previously published by the authors\footnotemark{} (Appendix A of \cite{Access_Taha_2024}), but now framed within a rigorous mathematical context.
Section 5 presents a case study of circulating currents in the windings of an electric machine.

\footnotetext{The property and its proof can be referred to in the preceding paper \cite{Access_Taha_2024}: \url{https://ieeexplore.ieee.org/document/10366258}.}

\section{Terminology and Nomenclature}

The basic terms related to windings, circulating currents, and losses used throughout this paper are listed below:
\begin{itemize}
    \item \textbf{Strand}: A single conductor with small dimensions (to mitigate both the skin and proximity effects).
    \item \textbf{Bundle}: A winding composed of several strands connected in parallel.
    \item \textbf{Bundle-level proximity effect losses} or \textbf{Circulating current losses}: Overall Losses occurring in windings due to the effect of circulating currents.
    \item \textbf{Period $T$}: Period of the AC electrical quantities.
    \item \textbf{Current $I(t)$}: Total current flowing in the bundle at the instant $t$.
    \item \textbf{Current $I_i (t)$}: Current flowing in the strand $i$ at the instant $t$ considering the circulating currents.
    \item \textbf{Current $I_{CC=0,i} (t)$}: Current flowing in the strand $i$ at the instant $t$ considering no circulating currents occurring.
    \item \textbf{Current $I_{RMS}$}: RMS value of the total current flowing in the bundle:
    \begin{equation}
    I_{RMS} = \sqrt{\frac{1}{T} \int_{0}^{T}I^2(t) \,dt} \\
\label{RMSexpression1}
\end{equation}
    \item \textbf{Current $I_{RMS,i}$}: RMS value of the current flowing in the strand $i$ considering the circulating currents:
    \begin{equation}
    I_{RMS,i} = \sqrt{\frac{1}{T} \int_{0}^{T}I_i^2(t) \,dt} \\
\label{RMSexpression2}
\end{equation}
    \item \textbf{Current $I_{RMS,CC=0,i}$}: RMS value of the current flowing in the strand $i$ considering no circulating currents occurring:
    \begin{equation}
    I_{RMS,CC=0,i}= \sqrt{\frac{1}{T} \int_{0}^{T} I_{CC=0,i}^2(t) \,dt} \\
\label{RMSexpression3}
\end{equation}
    \item \textbf{Resistance $R_{DC}$}: DC resistance of the bundle (composed of $n$ parallel strands).
    \item \textbf{Resistance $R_{DC,i}$}: DC resistance of the strand $i$.
    \item \textbf{Losses $P_{CC=0,i}$}: DC Losses occuring in the strand $i$ considering no circulating currents:
    \begin{equation}
    P_{CC=0,i}= R_{DC,i} \times I_{RMS,CC=0,i}^2 \\
\end{equation}
    \item \textbf{Losses $P_{CC=0}$}: DC Losses occuring in the bundle considering no circulating currents:
    \begin{equation}
    P_{CC=0}= \sum_{i=1}^{n} P_{CC=0,i} \\
\end{equation}
    \item \textbf{Losses $P_{CC,i}$}: Losses occuring in the strand $i$ considering the circulating currents:
    \begin{equation}
    P_{CC,i}= R_{DC,i} \times I_{RMS,i}^2 \\
\end{equation}
    \item \textbf{Losses $P_{CC}$}: Losses occuring in the bundle considering the circulating currents (Bundle-level proximity effect losses):
    \begin{equation}
    P_{CC}= \sum_{i=1}^{n} P_{CC,i} \\
\end{equation}
\end{itemize}

\begin{figure}[htbp]
\begin{center}
\includegraphics[scale=0.67]{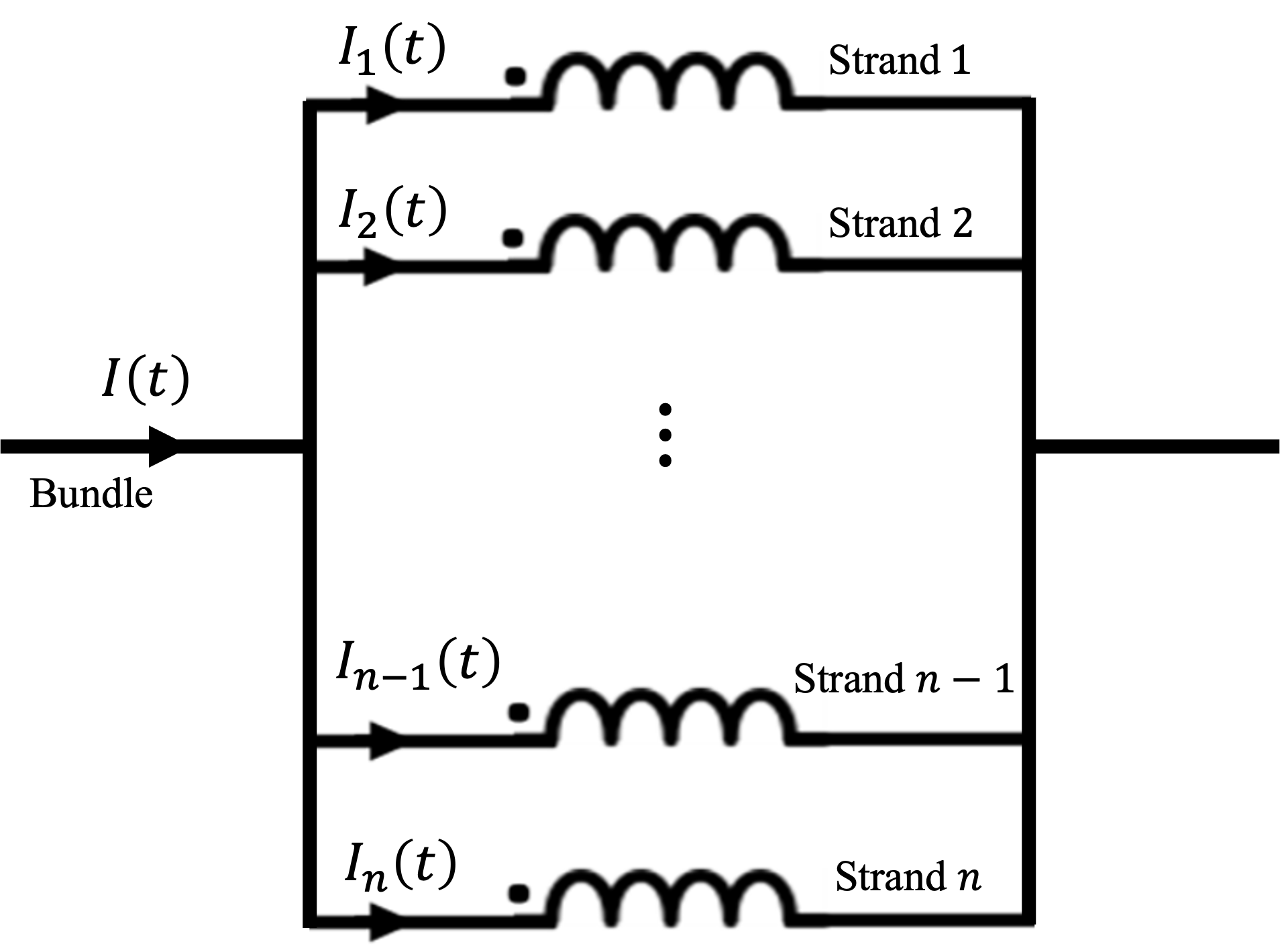}
\caption{Bundle composed of $n$ parallel strands}
\label{CC}
\end{center}
\end{figure}

The properties presented in this paper are applicable to both DC and AC cases. Without loss of generality, we are going to focus on the AC case in this work. Figure \ref{CC} shows the case of $n$ strands connected in parallel, all having the same dimensions and, consequently, the same DC resistance.

\section{Circulating Currents: Mathematical Definition}
Circulating currents can be defined mathematically as follows:

\begin{definition}
    Circulating currents \underline{occur} in a bundle, composed of $n$ parallel strands, if and only if:
    \begin{equation}
        \exists i \in \llbracket 1,n\rrbracket, \exists t \in [0,T],  I_i (t) \neq \frac{I(t)}{n}
    \label{definition1}
    \end{equation}
\end{definition}

The absence of circulating currents in a bundle can be correspondingly defined, by complement, as:
\begin{definition}
    Circulating currents \underline{do not occur} in a bundle, composed of $n$ parallel strands, if and only if:
    \begin{equation}
        \forall i \in \llbracket 1,n\rrbracket, \forall t \in [0,T],  I_i (t) = \frac{I(t)}{n}
    \label{definition2}
    \end{equation}
\end{definition}

Hence, when circulating currents do not occur in windings, the total current is evenly distributed among the $n$ parallel strands, a condition that also holds true under no load (zero total current flowing through the bundle), thereby:
\begin{equation}
        \forall i \in \llbracket 1,n\rrbracket, \forall t \in [0,T], I_{CC=0,i} (t) = \frac{I(t)}{n}
        \label{def_I_CC0_i}
\end{equation}
Evaluating currents flowing in each strand is essential to evaluate Bundle-level proximity effect losses. There exists various models in the literature, to evaluate these losses, which have been summarized and overviewed in \cite{Access_Taha_2024}.

\section{Fundamental Property and Proof}

The effect of circulating currents has a fundamental property that holds true across all applications (e.g., transformers, motors, generators, parallel circuits, etc.). This property and its proof are presented below.

\begin{property}\footnotemark{}

\begin{equation}
        \text{Circulating Currents } \Leftrightarrow  P_{CC=0} < P_{CC}
\label{property}
\end{equation}

\end{property}

\footnotetext{The property and its proof are presented here in a mathematically rigorous form and can be cited from the preceding paper \cite{Access_Taha_2024}: \url{https://ieeexplore.ieee.org/document/10366258}.}

There are two key points in this property:
\begin{itemize}
    \item $(\Rightarrow)$ When circulating currents occur in a bundle, the resulting losses are always higher than in the case with no circulating currents.
    \item $(\Leftarrow)$ The minimum achievable losses in a bundle are uniquely attained when the current is evenly distributed among the parallel strands.
\end{itemize}

\begin{proof}

We are going to prove both the direct and reciprocal implications of this property.

Let's consider a bundle composed of $n$ strands connected in parallel as shown in Figure \ref{CC}. The instantaneous total current flowing in the bundle is denoted $I(t)$ and the instantaneous current flowing in the strand $i$ is denoted $I_{i} (t)$. Since we are focusing on the AC case, we can express the current flowing in the strand $i$ as:
\begin{equation}
        \exists \alpha_i: [0,T] \xrightarrow{} \mathbf{R}, \forall t \in [0,T], I_i(t)=\alpha_i(t) \times I(t)
        \label{relstrdcurrent}
\end{equation}

The function $\alpha_i$ is not necessarily constant and can take both negative and positive values.

We have the following equality:
\begin{equation}
        \forall t\in[0,T], I(t)=\sum_{i=1}^{n} I_i(t)
\end{equation}
From which we can deduce the following equality:
\begin{equation}
        \forall t\in[0,T], 1 = \sum_{i=1}^{n} \alpha_i(t)
        \label{sumalpha}
\end{equation}

The direct implication of the property (\ref{property}) states that when circulating currents are occuring, the losses are higher than in the case with no circulating currents.

According to Cauchy-Schwartz inequality, for any vectors $\mathbf{a} = (a_1,a_2,...,a_n) \in \mathbf{R} ^n$ and $\mathbf{b}=(b_1,b_2,...,b_n) \in \mathbf{R} ^n$, we have:

\begin{equation}
        \left[\sum_{i=1}^{n} a_i \times b_i\right]^2 \le \left[\sum_{i=1}^{n} a_i^2\right] \times \left[\sum_{i=1}^{n} b_i^2\right]
        \label{Cauchy_Schwartz}
\end{equation}

Without the loss of generality, we can consider the vector $\mathbf{b} (t)$ variable with time. That is, at each instant $t$, we have a vector $\mathbf{b} (t)  \in \mathbf{R} ^n$. Let's consider the following vectors $\mathbf{a}$ and $\mathbf{b} (t)$:
\begin{equation}
\arraycolsep=1.4pt\def\arraystretch{1.25}
\begin{array}{ll}
    & \mathbf{a} = \left(a_1,a_2,...,a_n\right) = \left(1/n,1/n,...,1/n\right) \\
    & \mathbf{b} (t) = \left(b_1(t),b_2(t),...,b_n(t)\right) = \left(\alpha_1(t),\alpha_2(t),...,\alpha_n(t)\right)
\label{defab}
\end{array}
\end{equation}

Applying the Cauchy-Schwartz inequality (\ref{Cauchy_Schwartz}) yields:
\begin{equation}
        \left[\sum_{i=1}^{n} \frac{1}{n} \times \alpha_i(t)\right]^2 \le \left[\sum_{i=1}^{n} \left(\frac{1}{n}\right)^2\right] \times \left[\sum_{i=1}^{n} \alpha_i^2(t)\right]
\end{equation}
Hence,
\begin{equation}
        \frac{1}{n} \left[\sum_{i=1}^{n} \alpha_i(t)\right]^2 \le \sum_{i=1}^{n} \alpha_i^2(t)
\end{equation}

Which after using (\ref{sumalpha}) becomes:
\begin{equation}
        \frac{1}{n} \le \sum_{i=1}^{n} \alpha_i^2(t)
\end{equation}

Multiplying with $I^2(t) \ge 0 $ both sides yields:
\begin{equation}
        \frac{1}{n} I^2(t) \le \sum_{i=1}^{n} \alpha_i^2(t) \times I^2(t)
\end{equation}

Using (\ref{relstrdcurrent}) and reformulating both left and right sides gives:
\begin{equation}
        \sum_{i=1}^{n} \left(\frac{I(t)}{n}\right)^2 \le \sum_{i=1}^{n} I_i^2(t)
\end{equation}

Multiplying both sides with $R_{DC,i} \times \frac{1}{T}$ and integrating over the period $T$ results in:
\begin{equation}
        \sum_{i=1}^{n} R_{DC,i} \times \left[\frac{1}{T} \int_{0}^{T} \left(\frac{I(t)}{n}\right)^2 \,dt  \right] \le \sum_{i=1}^{n} R_{DC,i} \times \left[\frac{1}{T} \int_{0}^{T} I_i^2(t) \,dt \right]
        \label{eqlossdetailed1}
\end{equation}

Which, by applying (\ref{def_I_CC0_i}), can be rewritten as:
\begin{equation}
        \sum_{i=1}^{n} R_{DC,i} \times \left[\frac{1}{T} \int_{0}^{T} I_{CC=0,i}^2(t) \,dt  \right] \le \sum_{i=1}^{n} R_{DC,i} \times \left[\frac{1}{T} \int_{0}^{T} I_i^2(t) \,dt \right]
        \label{eqlossdetailed2}
\end{equation}

Substituting (\ref{RMSexpression2}) and (\ref{RMSexpression3}) in (\ref{eqlossdetailed2}) gives:

\begin{equation}
        \sum_{i=1}^{n} R_{DC,i} \times I_{RMS,CC=0,i}^2 \le \sum_{i=1}^{n} R_{DC,i} \times I_{RMS,i}^2
\end{equation}

Therefore,
\begin{equation}
        \sum_{i=1}^{n} P_{CC=0,i} \le \sum_{i=1}^{n} P_{CC,i}
\end{equation}

By summing all the losses in the strands, we obtain:
\begin{equation}
        P_{CC=0} \le P_{CC}
        \label{lossequation}
\end{equation}

Now, we need to deal with the case when the two terms are equal. The equality case in (\ref{lossequation}) is equivalent to the equality case in (\ref{Cauchy_Schwartz}).

Let's assume that we have the equality case in (\ref{lossequation}):
\begin{equation}
        P_{CC=0} = P_{CC}
        \label{lossequalequation}
\end{equation}
Therefore, we have also the equality in (\ref{Cauchy_Schwartz}):
\begin{equation}
        \left[\sum_{i=1}^{n} a_i \times b_i\right]^2 = \left[\sum_{i=1}^{n} a_i^2\right] \times \left[\sum_{i=1}^{n} b_i^2\right]
        \label{CSequal}
\end{equation}

According to Cauchy-Schwartz inequality, the equality occurs in (\ref{Cauchy_Schwartz}) if and only if the two vectors $\mathbf{a}$ and $\mathbf{b}$ are collinear, that is:
\begin{equation}
\begin{split}
        \exists \beta \in \mathbf{R}^*, \mathbf{a} = \beta. \mathbf{b}
\end{split}
\label{betaequ}
\end{equation}

Which, after substituting (\ref{defab}) in (\ref{betaequ}), yields:
 \begin{equation}
     \forall i \in \llbracket 1~;~ n \rrbracket, \frac{1}{n} = \beta \times \alpha_i(t)
     \label{equbetafori}
 \end{equation}

By summing both sides in (\ref{equbetafori}) for all $i \in \llbracket 1~;~ n \rrbracket$, we obtain:
\begin{equation}
\begin{split}
        \sum_{i=1}^{n} \frac{1}{n} = \sum_{i=1}^{n} \beta \times \alpha_i(t) = \beta \sum_{i=1}^{n} \alpha_i(t)
\end{split}
\label{equsumalpha1}
\end{equation}

Using (\ref{sumalpha}) yields:
\begin{equation}
\begin{split}
        1 = \beta
\end{split}
\end{equation}

This means that the vectors $\mathbf{a}$ and $\mathbf{b}$ are equal, therefore:
\begin{equation}
    \forall i \in \llbracket 1~;~ n \rrbracket, \forall t \in [0,T], \frac{1}{n}=\alpha_i(t)
\end{equation}

We conclude also that:
\begin{equation}
\begin{split}
    \forall i \in \llbracket 1~;~ n \rrbracket, \forall t \in [0,T], & I_i(t)=\alpha_i(t) \times I(t) \\
    & I_i(t) = \frac{I(t)}{n} = I_{CC=0,i}
\end{split}
\end{equation}

This means that the current is evenly shared between the $n$ parallel strands, in which case circulating currents do not occur in the windings, according to the definition (\ref{definition2}). Therefore, the equalities (\ref{lossequalequation}) and (\ref{CSequal}) are equivalent to no circulating currents occuring in the bundle. \\

\textbf{Summary:}

$(\Rightarrow)$ The inequality $P_{CC=0} \le P_{CC}$ (\ref{lossequation}) remains valid in all the cases. If circulating  currents occur, the equality $P_{CC=0}=P_{CC}$ (\ref{lossequalequation}) is not verified as shown above. Therefore, (\ref{lossequation}) becomes a strict inequality and the right side of the property (\ref{property}) is obtained. The direct implication of the property (\ref{property}) is then proven.

$(\Leftarrow)$The reciprocal implication is proved by contraposition. If no circulating currents occur, according to the result above, the equality $P_{CC=0}=P_{CC}$ (\ref{lossequalequation}) is verified. Since the inequality $P_{CC=0} \le P_{CC}$ (\ref{lossequation}) holds true in all the cases (whether circulating currents occur or no), then $P_{CC=0}=P_{CC}$ is equivalent to $P_{CC=0} \ge P_{CC}$ (because the inequality $P_{CC=0} > P_{CC}$ is never verified). Hence, no circulating currents implies $P_{CC=0} \ge P_{CC}$. Therefore, by contraposition, the reciprocal implication of the property (\ref{property}) is proven.

\end{proof}

\section{Case Application: Windings of Electric Machine}

\begin{figure}[!h]
\centering
\begin{subfigure}{.49\textwidth}
    \centering
    \includegraphics[clip,trim={0cm} {0cm} {1.71cm} {0.6cm},width=.9\linewidth]{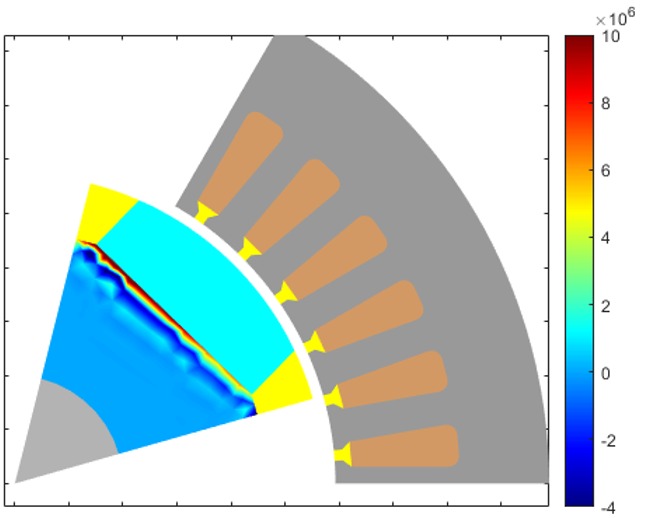}
    \caption{}
    \label{EM}
\end{subfigure}%
\begin{subfigure}{.5\textwidth}
    \centering
    \includegraphics[width=.8\textwidth]{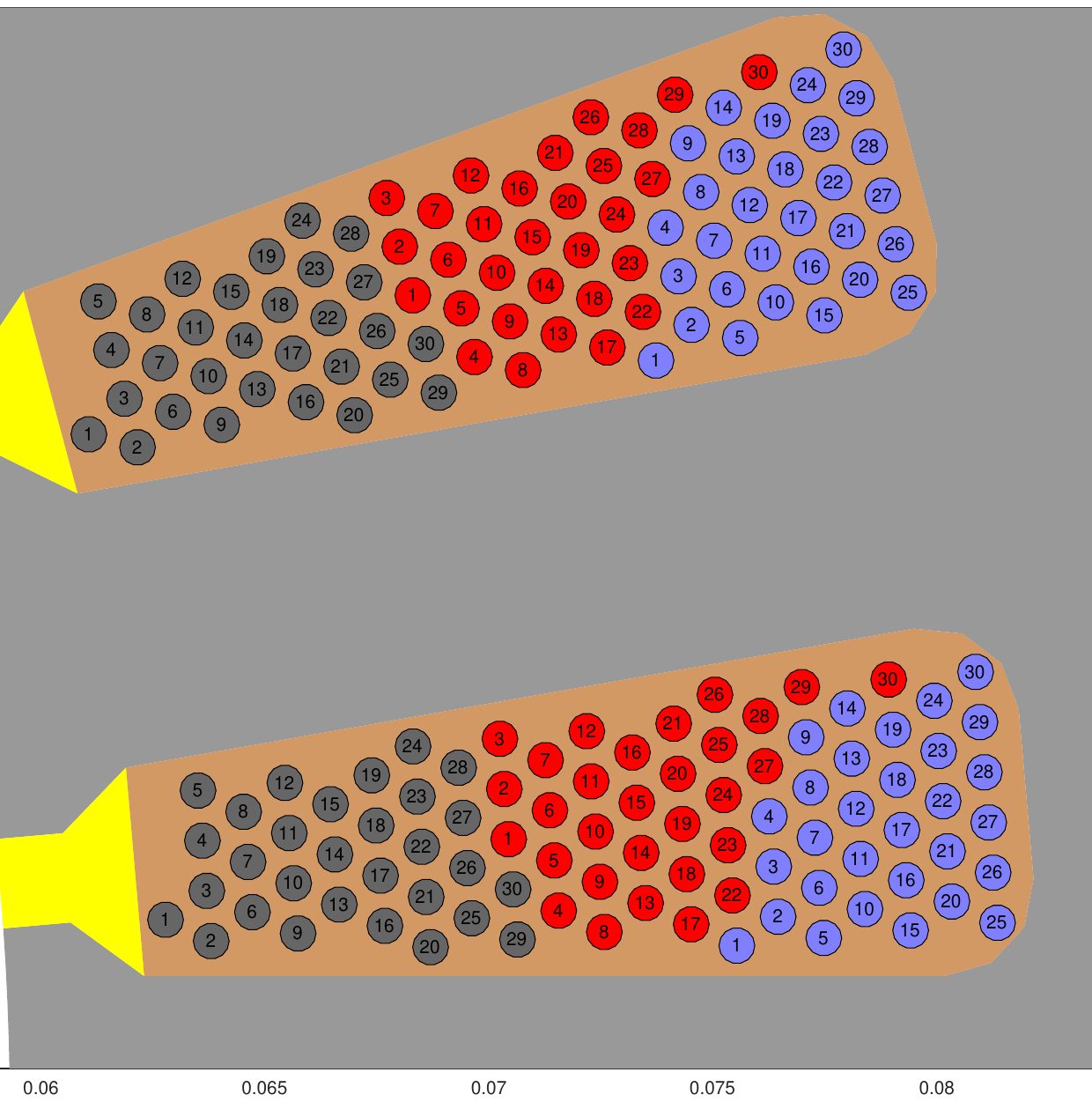}
    \vspace{-4.3em}
    \caption{}
    \label{SL}
\end{subfigure}
\caption{(a) Studied Machine (S-PMSM) and (b) Slot Layout with specified placement of the strands}
\end{figure}

The studied electric machine is a Surface Permanent Magnet Synchronous Machine (S-PMSM) as depicted in Figure \ref{EM}. The rotor is composed of $6$ pole and the stator is composed of $36$ slots and has a distributed winding, allocated to the three phases. Therefore, the stator has $2$ slots per pole and per phase. Each stator phase (A, B, C) is supplied with a pure sinusoidal current. The slot layout shown in Figure \ref{SL} illustrates the locations of the strands. Each phase consists of $30$ parallel strands, numbered from $1$ to $30$ in Figure \ref{SL}, and has $3$ turns per slot, represented in three different colors (blue, red, and black). The layout is identical for all three phases, with the strands occupying the same positions in the slots.

\vspace{-3em}

\begin{figure}[!ht]
\begin{center}
\includegraphics[width=0.7\textwidth]{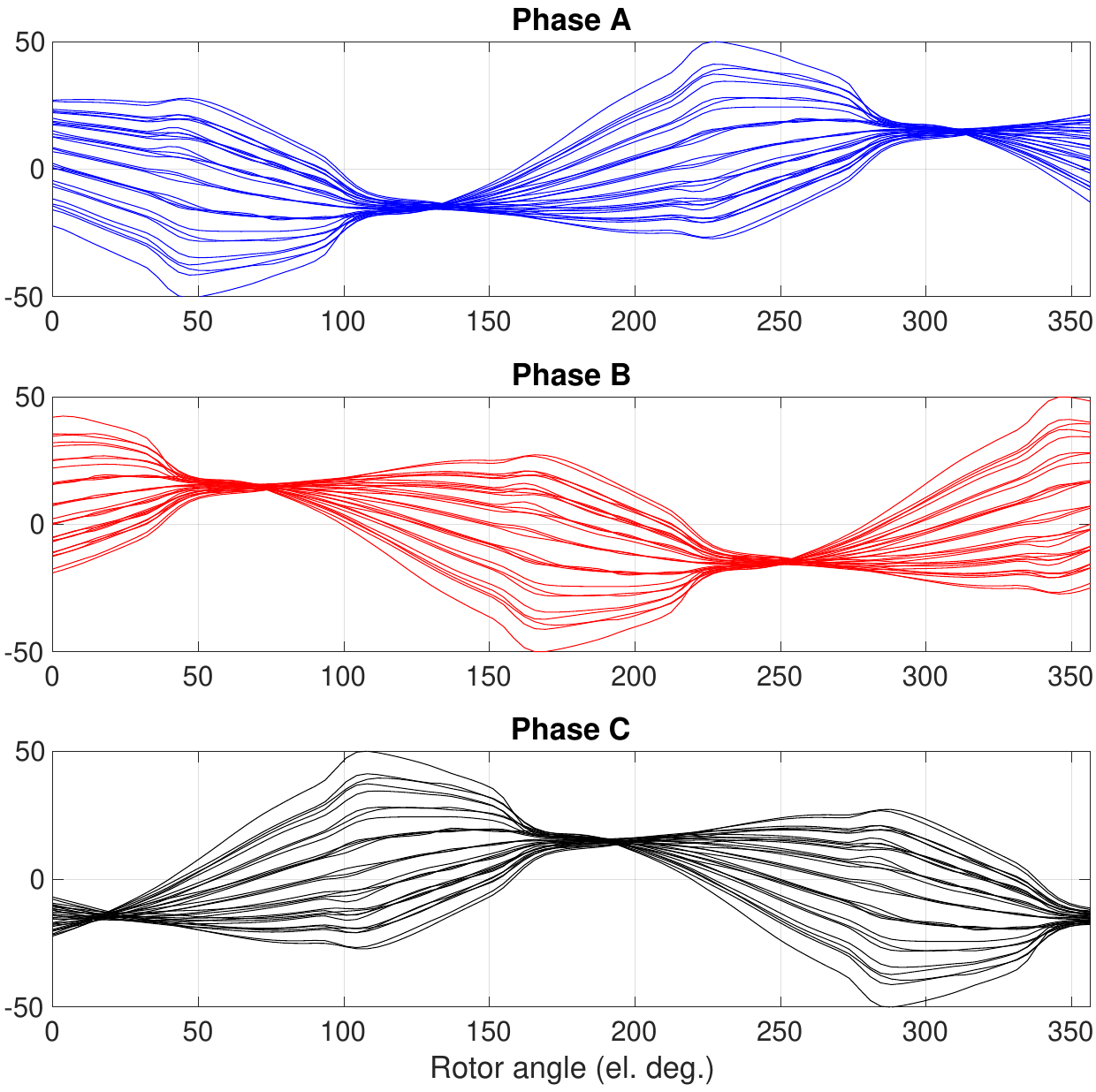}
\vspace{-6em}
\caption{Waveforms of the currents flowing in the parallel strands in each phase}
\label{CW}
\end{center}
\end{figure}

A co-simulation (finite element method coupled with circuit analysis) is conducted to evaluate the currents flowing through the 30 parallel strands. In the ideal case, the total current is evenly distributed among the 30 strands, resulting in equal magnitudes and phases for the currents in all parallel strands. Therefore, in the ideal case, the current flowing in each strand has a sinusoidal waveform similarly to that of the total current flowing in the bundle. However, due to variations in inductance and electric potential, circulating currents emerge, leading to an uneven distribution of current among the strands. Figure \ref{CW} illustrates the obtained results of the currents flowing in the strands of each phase. The currents flowing in the 30 strands in each phase do not exhibit sinusoidal waveforms, and they are neither equal in magnitude nor in phase, highlighting the effect of circulating currents. The behavior across the three phases is similar.

\vspace{-6em}

\vspace{-4em}
\section{Conclusion}

In this paper, we have defined the circulating currents in windings. This effect arises due to differences in inductance and electric potential among the parallel strands, leading to increased losses in the windings. This fundamental property, though intuitive, is widely acknowledged in the literature. We have provided a formal proof of this property governing circulating currents, extending the authors' previous work and framing it within a rigorous mathematical context. A case study of an electric machine was examined through co-simulation (finite element method coupled with circuit analysis) to evaluate the currents flowing through the parallel strands in each phase. The results demonstrated that circulating currents cause an uneven distribution, leading to non-sinusoidal waveforms and unequal magnitudes across the parallel strands, thereby resulting in higher losses. Although this paper focuses on an application in the windings of electric machines, circulating currents can occur in various other applications, and this fundamental property remains valid in all cases.

\section*{Acknowledgment}
This research is funded by the Research Council of Finland CoE HiECSs under Grant 346438.


\bibliographystyle{IEEEtran}

\bibliography{Document_CC}

\end{document}